\documentclass[fleqn,usenatbib]{mnras}

\usepackage{graphicx}
\usepackage{times}
\usepackage{grffile} %
\usepackage{physics}
\usepackage{amsmath}
\usepackage{amssymb}
\usepackage{cleveref}
\usepackage{xspace}
\usepackage{booktabs}
\usepackage{siunitx}
\DeclareSIUnit{\hred}{\mathit{h}}
\DeclareSIUnit{\Msun}{M_\odot}
\DeclareSIUnit\pc{pc}
\DeclareSIUnit\kpc{kpc}
\DeclareSIUnit\Mpc{Mpc}
\DeclareSIUnit\lightyear{ly}
\usepackage{xcolor}
\usepackage{soul}
\usepackage[normalem]{ulem}
\usepackage{enumerate}
\usepackage{tabularx}

\definecolor{bubbles}{rgb}{0.91, 1.0, 1.0}
\definecolor{aquamarine}{rgb}{0.5, 1.0, 0.83}
\definecolor{bubblegum}{rgb}{0.99, 0.76, 0.8}
\definecolor{bluebell}{rgb}{0.74, 0.74, 0.92}
\definecolor{dollarbill}{rgb}{0.72, 0.93, 0.6}

\crefname{table}{Table}{Tables}
\Crefname{table}{Table}{Tables}

\crefname{equation}{Eq.}{Eqs.}
\Crefname{equation}{Equation}{Equations}

\crefname{figure}{Fig.}{Figs.}
\Crefname{figure}{Figure}{Figures}

\crefname{section}{Section}{Sections}
\Crefname{section}{Section}{Sections}

\renewcommand{\qq}{\vec{q}}

\newcommand{\xx}{\vec{x}}

\newcommand{\ie}{i.e.\xspace}
\newcommand{\eg}{e.g.\xspace}

\newcommand{\ramses}{\textsc{Ramses}\xspace}

\newcommand{\genetic}{\textsc{genetIC}\xspace}
\newcommand{\adaptahop}{\textsc{AdaptaHOP}\xspace}

\newcommand{\AM}{angular momentum\xspace}
\newcommand{\spAM}{specific angular momentum\xspace}

\newcommand{\TTT}{tidal torque theory\xspace}
\renewcommand{\vec}[1]{\vb*{#1}}

\sethlcolor{bluebell}

\title[Angular momentum evolution can be predicted from cosmological initial conditions]{%
   Angular momentum evolution can be predicted from\\ cosmological initial conditions%
}

\author[C.~Cadiou et al.]{Corentin~Cadiou$^{1}$\thanks{c.cadiou@ucl.ac.uk},
Andrew~Pontzen$^{1}$,
Hiranya~V.~Peiris$^{1,2}$
\\
$^{1}$Department of Physics and Astronomy, University College London, Gower Street, London WC1E 6BT, United-Kingdom\\
$^{2}$The Oskar Klein Centre for Cosmoparticle Physics, Department of Physics, Stockholm University, AlbaNova, Stockholm SE-106 91, Sweden
}

\date{Accepted 2021 February 3. Received 2021 January 30; in original form 2020 December 14.}
\pubyear{2020}

\begin{document}
\label{firstpage}
\pagerange{\pageref{firstpage}--\pageref{lastpage}}
\maketitle

\begin{abstract}
   The angular momentum of dark matter haloes controls their spin magnitude and orientation, which in turn influences the galaxies therein.
   However, the process by which dark matter haloes acquire angular momentum is not fully understood; in particular, it is unclear whether angular momentum growth is stochastic.
   To address this question, we extend the genetic modification  technique to allow control over the angular momentum of any region in the initial conditions. Using this technique to produce a sequence of modified simulations, we can then investigate whether changes to the angular momentum of a specified region in the evolved universe can be accurately predicted from changes in the initial conditions alone. 
   We find that the angular momentum in regions with modified initial conditions can be predicted between 2 and 4 times more accurately than expected from applying \TTT.
   This result is masked when analysing the angular momentum of haloes, because particles in the outskirts of haloes dominate the angular momentum budget.
   We conclude that the angular momentum of Lagrangian patches is highly predictable from the initial conditions, with apparent chaotic behaviour being driven by stochastic changes to the arbitrary boundary defining the halo.
\end{abstract}


\section{Introduction}
A pressing question in galaxy formation theory is the origin of the angular momentum of dark matter haloes and of their host galaxies.
Angular momentum controls the spin magnitude and orientation of dark matter haloes, which in turn influences the galaxies they host \citep{fall_FormationRotationDisc_1980,mo_FormationGalacticDiscs_1998,bullock_UniversalAngularMomentum_2001,burkert_AngularMomentumDistribution_2016}.
Beyond its relevance to galaxy formation, the origin of angular momentum also has consequences for cosmology.
Due to its common cosmological origin, the angular momentum accreted by neighbouring galaxies tends to be aligned, which in turn contributes to the alignment of galaxy spins.
This effect is degenerate with the lensing signal, and therefore needs to be mitigated to extract cosmological information from weak lensing surveys (\eg \emph{Euclid}, \citealt{laureijs_EuclidDefinitionStudy_2011}; LSST, \citealt{2019ApJ...873..111I}).

The current understanding of the origin of the angular momentum of galaxies and dark matter haloes is built on the work of \cite{doroshkevich_SpaceStructurePerturbations_1970}, who derived the first models to predict the evolution of angular momentum within non-spherical regions.
The authors showed that angular momentum grows linearly under the Zel'dovich approximation \citep{zeldovich_GravitationalInstabilityApproximate_1970a}, while its magnitude in spherical regions grows only at second order \citep{peebles_OriginAngularMomentum_1969}.
The model was later improved by \cite{white_AngularMomentumGrowth_1984} to formulate `\TTT'.
The theory states that in the linear regime of structure formation, angular momentum is generated from slight misalignments of the inertia tensor of the Lagrangian patch with the tidal shear tensor.
As the proto-halo decouples from the expansion of the Universe and collapses, the coupling weakens and eventually vanishes.
{After} decoupling, \TTT assumes little evolution of the angular momentum.
Later studies confirmed that \TTT predictions are accurate within a factor of two if the angular momentum growth is stopped around turn-around \citep{sugerman_TestingLineartheoryPredictions_2000,porciani_TestingTidaltorqueTheory_2002a}.
Tidal torque theory provides a global explanation to the origin of angular momentum and successfully explains a variety of cosmological effects.
It was, for example, recently extended by \cite{codis_SpinAlignmentsCosmic_2015} to relate the structure of the tides imposed by the large-scale environment to the angular momentum acquisition of forming haloes, which provided a convincing explanation for the observed spin alignment signal \citep{tempel_EvidenceSpinAlignment_2013}.
Its predictions are, however, only accurate statistically {and display significant deviations when looking at individual halos} \citep{porciani_TestingTidaltorqueTheory_2002b}.
A different picture emerges in studies which focus on dark matter haloes instead of Lagrangian patches.
On average, the angular momentum of haloes grows with each merger \citep{vitvitska_OriginAngularMomentum_2002,peirani_AngularMomentumDark_2004,hetznecker_EvolutionDarkHalo_2006}.
This motivated the development of semi-analytical models, where the \AM evolution is controlled by the haloes' merger tree.
In these models, each dark matter halo is initially seeded with a random angular momentum.
After each merger, the angular momentum of the new halo is set by the \AM of its progenitors and their (randomly-drawn) orbital parameters \citep{vitvitska_OriginAngularMomentum_2002}.
Semi-analytical models were later refined using more accurate merger trees, halo concentrations and distributions of orbital parameters \citep{benson_GalaxyFormationSpanning_2010}.
{This allows accurate reproduction of the distribution}
of spin parameters measured in $N$-body simulations \citep{benson_RandomwalkModelDark_2020}.

The two competing approaches seem to be in tension: according to \TTT, the angular momentum of a given proto-halo can be derived from a linear analysis of its initial conditions.
On the other hand, the angular momentum yielded by semi-analytical models is intrinsically stochastic, and {depend on} the initial conditions indirectly, through the merger history.

In this paper, we address the question of whether the actual angular momentum of given regions in the Universe results from chaotic processes, or whether it can be predicted accurately from an analysis of the initial conditions.
To do so, we develop an extension to the genetic modification technique \citep{roth_GeneticallyModifiedhaloes_2016,stopyra_GenetICNewInitial_2020} to control the angular momentum of arbitrary regions in the initial conditions.
{The technique allows us to change the initial angular momentum through small modifications of the initial density field that minimally affect the large-scale environment.}
Using a suite of simulations, we then investigate how the angular momentum in the evolved Universe changes with modifications of the initial conditions.
{%
   We can thus test cause and effect relations between changes in the initial angular momentum and its late time evolution. This allows us to disentangle the effect of chaotic evolution from the effect of particle membership on angular momentum evolution.
}
In \cref{sec:methods}, we describe our simulation setup. Our results are presented in \cref{sec:results}, and we conclude in \cref{sec:discussion}.

\begin{table}
   \caption{Properties of the reference haloes at $z=0$. From left to right: the halo id, the virial mass and radius, the specific angular momentum and the spin parameter.}%
   \label{tab:halo_description}
   \centering
   \begin{tabularx}{\columnwidth}{cScSc}
      \toprule
      Halo & {$M_\mathrm{vir}/\SI{e12}{\Msun}$} & {$R_\mathrm{vir}/\si{\kpc}$} & {$l/ \SI{100}{\km.\s^{-1}.\kpc}$} & {Spin $\lambda$} \\
      \midrule
      123 &  5.3 & 450 & 50 & 0.024   \\
      189 & 16.8 & 660 & 32 & 0.006  \\
      207 & 15.4 & 640 & 75 & 0.013  \\
      246 & 24.2 & 750 & 209 & 0.033 \\
      374 &  2.8 & 370 & 593 & 0.120  \\
      386 &  6.3 & 480 & 178 & 0.053  \\
      390 &  7.5 & 510 & 109 & 0.014  \\
      \bottomrule
   \end{tabularx}
\end{table}

\begin{figure}
   \includegraphics[width=\columnwidth]{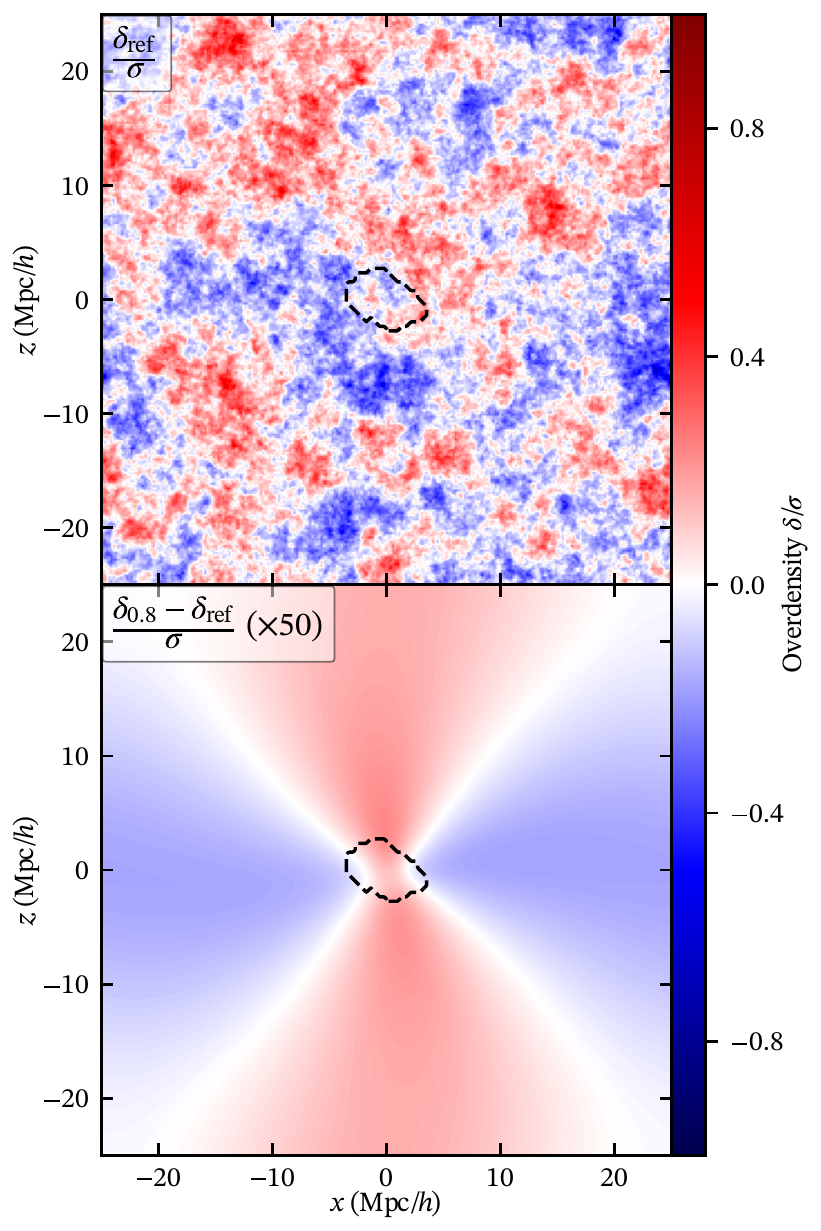}
   \caption{
      (\emph{Top}) Projected mean overdensity field in the reference simulation and
      (\emph{bottom}) amplitude of the overdensity modification of halo 386, for a change of its \AM by $\times 0.8$.
      Both panels show values in the initial conditions, in units of the overdensity standard deviation.
      The Lagrangian region is shown in the centre (black dashed contours).
   }%
   \label{fig:initial_constraint}
\end{figure}
\section{Methods}
\label{sec:methods}

All our simulations employ the code \ramses{} \citep{teyssier_CosmologicalHydrodynamicsAdaptive_2002a}.
We first perform a reference dark-matter-only cosmological $N$-body simulation.
The adopted cosmology has a total matter density $\Omega_\mathrm{m} = 0.31$, a dark energy density of $\Omega_\Lambda = 0.69$, a Hubble constant of $H_0 = \SI{67.3}{km.s^{-1}.Mpc^{-1}}$, a variance at \SI{8}{Mpc\per\hred} $\sigma_8 = 0.8159$ and a non-linear power spectrum index of $n_\mathrm{s} = 0.9667$, compatible with the parameters of \cite{planckcollaboration_Planck2015Results_2015}.
We start with a coarse grid of $256^3$ (level 8) in a box of comoving size \SI{50}{\Mpc\per\hred} for a dark matter particle mass of \SI{9e8}{\Msun}.
We use a quasi-Lagrangian refinement strategy: a cell is refined if it contains 8 or more dark matter particles.
The gravitational potential is computed on the adaptive grid with a multigrid particle-mesh solver up to level 11 \citep{guillet_SimpleMultigridScheme_2011} and with the default conjugate gradient method for finer levels. We have an effective minimum softening length of physical size \SI{0.8}{kpc} imposed by the grid resolution.
We dump a snapshot every \SI{250}{Myr} for a total of $55$ outputs.

We extract a halo catalogue using \adaptahop{} \citep{aubert_OriginImplicationsDark_2004}, with the `Most massive Sub-node Method' and the parameters proposed by \cite{tweed_BuildingMergerTrees_2009a}.
Only haloes with 200 or more particles are kept. {We verified that performing the same analysis with halos containing more than {\num{2000}} particles yield similar results, although with smaller statistics.}

We then select the first 7 haloes from the halo catalogue that are more massive than \SI{5e12}{\Msun} ({{\SI{3.5e12}{\Msun\per\hred}},} more than \num{5000} particles) at $z=0$, and whose Lagrangian patch does not contain any particle within \SI{0.1}{{\emph{L}}_\mathrm{box}} of the box sides\footnote{This simplifies the analysis, as we can ignore periodicity when computing distances within each Lagrangian patch.}.
The main properties of the selected haloes are presented in \cref{tab:halo_description}. Here, we used the following definition to compute the spin parameter: $\lambda \equiv l\sqrt{K+U}/ GM^{3/2}$, %
{where $G$ is the gravitational constant and $l,K,U$ and $M$ are respectively the specific angular momentum, the kinetic energy, the potential energy and the halo mass as computed by the halo finder.}

\begin{figure*}
   \includegraphics[width=\textwidth]{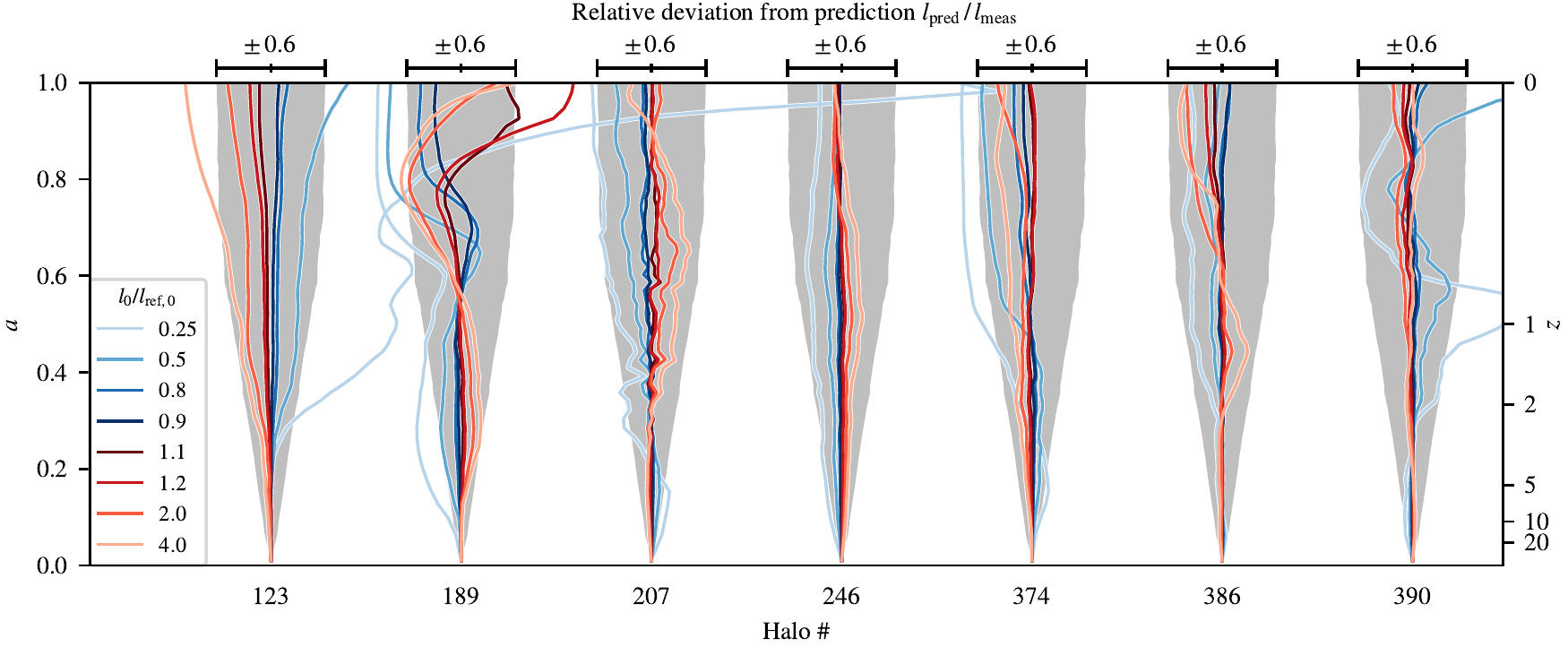}
   \caption{
      Deviations of the predicted \spAM to the measured \spAM, $l_\mathrm{pred}/l_\mathrm{meas}-1$, in the Lagrangian patch of all seven haloes.
      The deviations are shown for different initial \spAM modifications, $l_0/l_{\mathrm{ref},0}$.
      We also show for comparison the amplitude of the \SI{68}{\percent} confidence interval of the deviations from \TTT's predictions, $l_\mathrm{pred,TTT}/l_\mathrm{meas}-1$ (gray region).
      Most resimulations have their ratio well within this confidence interval.
      This shows that the \spAM{} in resimulations can be predicted from the reference simulation more accurately than using \TTT.
   }
   \label{fig:spAM-multiple-haloes-with-TTT-large-resim}
\end{figure*}
For each halo, we trace back the region that contains all particles within the halo at $z=0$ (their Lagrangian patch) to the initial conditions at $z=99$ .
We then generate, for each region, a set of eight genetically modified initial conditions with a modified \AM: $\times 1/4, \times 1/2, \times 0.8, \times 0.9, \times 1.1, \times 1.2, \times 2$ and $\times 4$.
We generate the initial conditions using the code \genetic v1.2 \citep{roth_GeneticallyModifiedhaloes_2016,rey_QuadraticGeneticModifications_2018,stopyra_GenetICNewInitial_2020,andrew_pontzen_2021_4509596}, that includes \AM genetic modifications; see Appendix~\ref{sec:details-genetic-implementation} for the details of the implementation.
{%
   This technique allows us to arbitrarily modify the total angular momentum within the patch by altering the initial density field.
   The genetic modification method ensures that these modifications are minimal and keep the large-scale environment as near unchanged as possible.
   This provides us with a numerical setup to quantify how changes in the initial angular momentum affect its growth rate and test whether the late-time evolution is chaotic or not.
}%
For each patch, we modify the three components of the initial \AM by the same relative amount, while keeping the mean density fixed.
In the remainder of the paper, we will study the evolution of the \AM within these regions.

The result of the $\times 0.8$ modification for the initial conditions of halo 386 can be seen in \cref{fig:initial_constraint}.
We show in the top panel the mean density field and in the bottom panel the density modification required to alter the \AM.
The figure shows that the modification extends significantly outside the Lagrangian region, and displays a distinct quadrupolar structure.
This can be understood qualitatively by recalling that, under the Zel'dovich approximation, the angular momentum of a region originates from peculiar {velocity gradients} (as detailed in Appendix~\ref{sec:details-TTT} {and underlined by \protect\citealt{neyrinck_HaloSpinPrimordial_2020}}).
{\cite{pontzen_EDGENewApproach_2020} showed that the density modifications corresponding to velocity modifications} extend to scales of the order of \SI{60}{Mpc\per\hred} in a $\Lambda$CDM universe.\footnote{We expect \AM modifications to translate into quadrupolar density modification{s}. \cite{pontzen_EDGENewApproach_2020} showed that velocity modifications translate into dipolar modifications of the density field, so that {the modification of the gradient of the peculiar velocity} required to modify the \AM translates into {a} quadrupolar density modification.
}
Since our simulations have a comoving size of \SI{50}{Mpc\per\hred}, we {therefore} expect the density modification to extend to the entire simulation domain.

We perform resimulations starting from each of the modified initial conditions ($7\times 8$ in total) with the same parameters as in the reference simulation.
For all resimulations, we compute the \AM of the region that was modified, and obtain its \spAM by dividing by the mass it contains (see \cref{tab:halo_description}).
\section{Results}
\label{sec:results}

Tidal torque theory provides a prediction for the time evolution of the \spAM in a region,
\begin{equation}
   l_\mathrm{pred,TTT}(z) = l_0 \times \frac{a^2(z)\dot{D}(z)}{a^2_0\dot{D}_0},
   \label{eq:predictor-TTT}
\end{equation}
where $l_0$ is the initial \spAM within the region, $a$ is the expansion factor, $D$ is the linear matter growth function, and $a_0$ and $D_0$ are their respective initial value.
Here, dots indicate time derivatives.
However, when performing modified simulations, we do not need to rely directly on \TTT.
Instead, we can use the measured value of the \spAM in the reference simulation to estimate the growth rate,
\begin{equation}
   l_\mathrm{pred}(z) = l_0 \times \frac{l_\mathrm{ref}(z)}{l_{\mathrm{ref},0}},
   \label{eq:predictor}
\end{equation}
where $l_\mathrm{ref}(z)$ is the measured magnitude of the \spAM in the reference simulation and $l_{\mathrm{ref},0}$ is its value in the initial conditions.

In our suite of resimulations, the ratio of the initial \spAM{}s, $l_0/l_{\mathrm{ref},0}$, is equal to the amplitude of the \spAM modification.
Using measurements of $l_\mathrm{ref}(z)$ and the $7\times 8$ modified resimulations with known $l_0/l_{\mathrm{ref},0}$, we can compare the accuracy of the ansatz \eqref{eq:predictor} to the accuracy of \TTT.
To do so, we show in \cref{fig:spAM-multiple-haloes-with-TTT-large-resim} the deviation from unity of the ratio of our prediction to the measured value, $l_\mathrm{pred}/l_\mathrm{meas}-1$ (as coloured lines, for different amplitudes of the modifications) for the 8 resimulations of the 7 haloes.
If the \spAM in the resimulations is proportional to its value in the reference simulation, the ratio will be equal to one.
In order to qualitatively compare any deviations from one to the accuracy of \TTT, we measure the relative \SI{68}{\percent} confidence interval of \TTT's predictions compared to the measured values\footnote{We define the relative confidence interval as the measure of the deviations around the median value, and is given by the ratio of the \SI{68}{\percent} confidence interval to the median of $l_{\mathrm{pred,TTT}}/l_\mathrm{meas}-1$. This allows us to correct for the fact that, by $z=0$, the predictions are systematically $\sim 3$ larger than the actual value.}, using all 700 haloes more massive than \SI{e12}{\Msun} in the reference simulation (gray bands).
Both \TTT and our ansatz \eqref{eq:predictor} are accurate in the early, linear Universe.
With increasing cosmic time, we observe that the accuracy of \TTT worsens, as shown by the widening of the \SI{68}{\percent} interval
{which, at $z=0$, has a width of $\pm 0.6$.}
The deviations of our ansatz from the measured value are statistically within the contours set by \TTT down to $z=0$, indicating that the ansatz \eqref{eq:predictor} is more accurate than \TTT.
We also note that large deviations are only ever obtained from large modifications of the initial \spAM (such as halo 123$\times 1/4$ and halo 189$\times 4$).

\begin{figure}
   \includegraphics[width=\columnwidth]{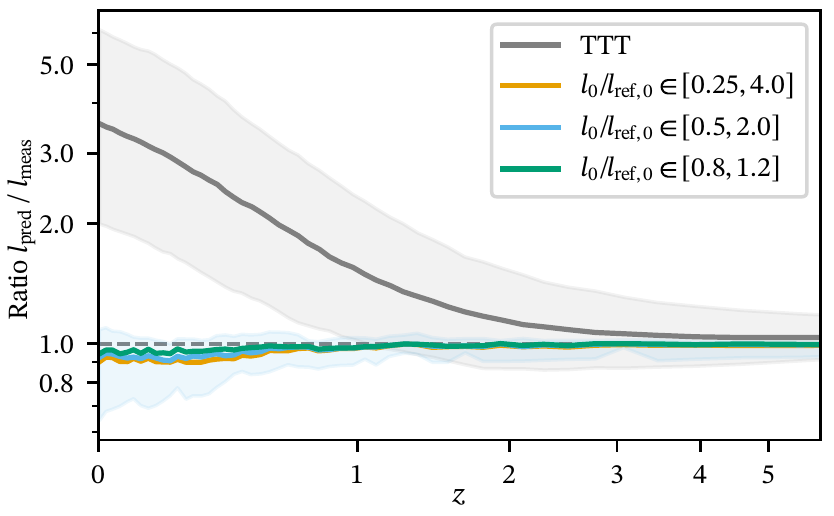}
   \caption{
      Evolution of the deviation of the predicted \AM from the measured value.
      Predictions rely either on \TTT (gray line) or on the reference simulation (from green to blue to orange in increasing order of the amplitude of modifications, as labelled).
      Shaded regions indicate \SI{68}{\percent} confidence intervals.
     Tidal torque theory systematically overpredicts the measured \spAM, and its predictions are significantly more scattered than predictions based on a reference simulation.
      The latter are consistent with unity: they are less biased and more accurate, especially at low redshift.
   }
   \label{fig:spAM-vs-TTT:median_and_scatter}
\end{figure}

\begin{figure}
   \includegraphics[width=\columnwidth]{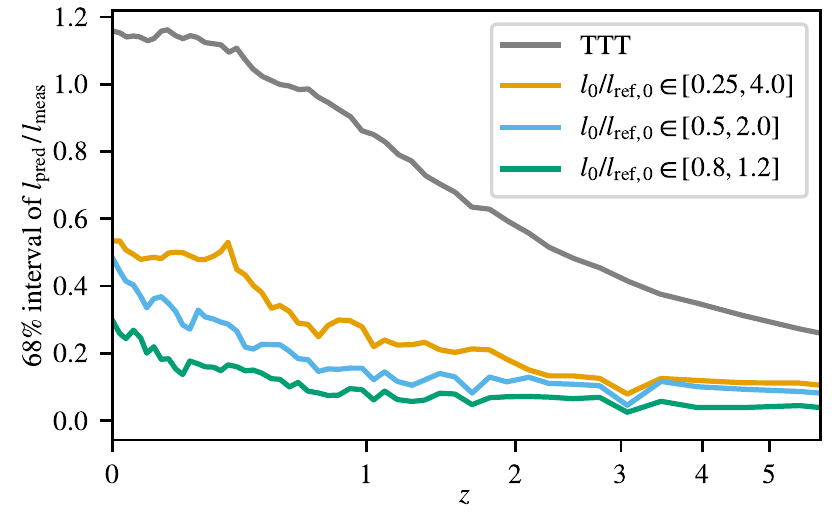}
   \caption{
      Time evolution of the \SI{68}{\percent} confidence interval of the ratio $l_\mathrm{pred}/l_\mathrm{meas}-1$, with the same colours as \cref{fig:spAM-vs-TTT:median_and_scatter}.
      At all times, predictions based on the reference simulation are less scattered around the true value than predictions from \TTT.
      Smaller modifications of the initial \spAM are progressively more accurate.
   }
   \label{fig:spAM-vs-TTT:scatter-evolution}
\end{figure}

Let us now study in more detail the evolution of the deviation of our predictions from the actual value with cosmic time.
We show in \cref{fig:spAM-vs-TTT:median_and_scatter} the evolution of the median (solid lines) and \SI{68}{\percent} confidence interval (shaded regions) of $l_\mathrm{pred}/l_\mathrm{meas}-1$ .
We compute the median and confidence interval of the ratio from three different subsets of resimulations: all of them ($0.25 \leq l_0/l_{\mathrm{ref},0} \leq 4$; orange line), all but most extreme modifications ($0.5 \leq l_0/l_{\mathrm{ref},0} \leq 2$; blue line and blue shaded area), and only those with small \spAM modifications ($0.8 \leq l_0/l_{\mathrm{ref},0} \leq 1.2$; bluish green line).
We show the confidence interval for one subset of resimulations for the sake of clarity, the two others being presented in \cref{fig:spAM-vs-TTT:scatter-evolution}.
We also show the median and \SI{68}{\percent} interval we obtain with \TTT (in gray).
The predictions made with ansatz \eqref{eq:predictor} slightly underestimate the actual \AM, as shown in \cref{fig:spAM-vs-TTT:median_and_scatter} by a weak deviation of the median from unity (of the order of $0.9-0.95$) at $z\lesssim 1$.
The \SI{68}{\percent} confidence interval however remains consistent with unity at all times.
Predictions from \TTT are also consistent with the measured \AM at early times $z\gtrapprox 1-2$, while at later times, we recover the well-established result that \AM growth stalls, and as a result \TTT starts overpredicting the actual \AM.
By $z=0$, \TTT overpredicts the \AM by a factor as large as $3$, in agreement with previous findings \citep{porciani_TestingTidaltorqueTheory_2002a}.
We conclude that the predictions from ansatz \eqref{eq:predictor} are less biased than those from \TTT, and remain consistent with the measured \AM down to $z=0$.
{We have also checked that similar conclusions can be made about the orientation of the angular momentum vector: ansatz \eqref{eq:predictor} provides a more accurate prediction of the angular momentum orientation than tidal torque theory.}

In order to compare the relative accuracy of \TTT predictions to those of ansatz \eqref{eq:predictor}, we show the evolution of the relative \SI{68}{\percent} confidence intervals of $l_\mathrm{pred}/l_\mathrm{meas}-1$ in \cref{fig:spAM-vs-TTT:scatter-evolution}.
We recover here the result that, at $z=0$, predictions from \TTT are only accurate within a factor of $\sim 2$ (fluctuations of the order of \SI{120}{\percent}).
By contrast, we observe that the predictions of ansatz \eqref{eq:predictor} are accurate within {$\sim 0.5$} of the measured \AM (for the largest modifications, $l_0/l_{\mathrm{ref},0}\in[0.25,4]$) and within {$\sim 0.3$} (for the smallest modifications, $l_0/l_{\mathrm{ref},0}\in[0.8,1.2]$) at $z=0$.
We also observe that predictions based on our ansatz remain accurate down to a later time compared to predictions from \TTT.
For example, the relative confidence interval reaches half its final value as early as $z\approx 2$ for \TTT, compared to $z\approx 1$ with our ansatz.

\section{Discussion}
\label{sec:discussion}

Our results demonstrate that the \AM of Lagrangian patches can be accurately predicted from the initial conditions, and allow us to rule out scenarios where it would be driven by chaotic or stochastic processes.
Using the ansatz \eqref{eq:predictor} we have shown that, for individual Lagrangian patches, the \AM can be predicted from the product of its initial value and a growth rate. {We found the accuracy of our predictions at $z=0$ to be between 2 and 4 times  better than the accuracy of tidal torque theory, with smaller modifications leading to more accurate predictions.
}

Given our findings, we argue that the problem of the origin of the \AM of dark matter haloes can be decomposed into two parts.
The first issue is that of the origin of \AM growth, which we addressed in this work.
Our results show that the growth rate of individual regions in resimulations can be estimated accurately from its measured value in a reference simulation.
We interpret this as evidence that the growth rate is an intrinsic property of a given region and its environment, and that \AM does not result from chaotic  processes.
One implication for \TTT is that its accuracy could be improved by predicting (rather than measuring) this growth rate from an analysis of the Lagrangian patch in the linear universe.

The second aspect is the problem of the initial value of the \AM in the linear Universe, \ie{} the term $l_0$ in \cref{eq:predictor-TTT,eq:predictor}.
As was already pointed out by~\cite{white_AngularMomentumGrowth_1984,porciani_TestingTidaltorqueTheory_2002b}, we would need a robust method to draw the boundaries of the Lagrangian patch to solve this problem. {For dark matter haloes, these two aspects are not independent because the tidal field, which controls the torquing, and the inertia tensor, which controls the shape of the Lagrangian patch, are correlated.
}%
This remains a topic for future research, and currently limits the possibility of using genetic modifications to control the \AM of haloes {rather than fixed regions} as modifications by construction deform the Lagrangian patch.
However, we argue that it should be possible to control the \AM of simulated galaxies.
The distribution of the angular momentum of the gas is much more concentrated, and more coherent radially than the dark matter \AM \citep{danovich_FourPhasesAngularmomentum_2015}.
We therefore expect the gas \AM to be less sensitive to deformation of its Lagrangian patch.
In particular, galaxies that are fed through cold flows are good candidates for testing this hypothesis.
The flows are aligned with the cosmic web which sets their angular momentum and direction \citep{danovich_CoplanarStreamsPancakes_2012,pichon_WhyGalacticSpins_2014}, and they feed most of the baryonic angular momentum to the galaxies \citep{birnboim_VirialShocksGalactic_2003,stewart_HighAngularMomentum_2017}.
We expect the structure of the cosmic web to be only marginally impacted by genetic modifications \citep{stopyra_GenetICNewInitial_2020}, giving us a potential lever for controlling the \AM in the flows, and hence in the infalling gas.
In the future, we will apply genetic modifications to study of the origin of \AM in galaxies, and how different \AM acquisition histories lead to different galactic properties, such as their morphology or disk structure.

\section*{Acknowledgements}

The authors acknowledge Nina Roth's unpublished preliminary work on genetic modifications of angular momentum, which motivated this project.
{We would like to thank R.~Teyssier and the anonymous referee for useful comments.}
This early work was supported by European Research Council under the European Community’s Seventh Framework Programme (FP7/2007-2013) / ERC grant agreement No.\@ 306478 CosmicDawn.
This project has received funding from the European Union's Horizon 2020 research and innovation programme under grant agreement No.\@ 818085 GMGalaxies.
HVP's work was partially supported by the research project grant `Understanding the Dynamic Universe' funded by the Knut and Alice Wallenberg Foundation under Dnr KAW 2018.0067.
AP was supported by the Royal Society.
{This work used computing equipment funded by the Research Capital Investment Fund (RCIF) provided by UKRI, and partially funded by the UCL Cosmoparticle Initiative.}
The analysis was carried out using \textsc{Python},
\textsc{Matplotlib} \citep{hunter2007matplotlib},
\textsc{Numpy} \citep{harris_ArrayProgrammingNumPy_2020},
\textsc{Pynbody} \citep{pontzen_PynbodyNBodySPH_2013} and
\textsc{Yt} \citep{turk_YtMulticodeAnalysis_2011}.

\section*{Data availability}
The data underlying this article will be shared on reasonable request to the corresponding author.

\bibliographystyle{mnras}
\bibliography{authors}

\appendix

\section{Genetic modifications of the angular momentum}
\label{sec:details-genetic-implementation}

{
   We provide in this section a qualitative description of the implementation of angular momentum genetic modifications.
   We refer to \cite{roth_GeneticallyModifiedhaloes_2016,rey_QuadraticGeneticModifications_2018,stopyra_GenetICNewInitial_2020} for a more formal derivation of the genetic modification method.
}

{
   Let us suppose we have initial conditions with values on a regular grid $f^{(i)}$, where $i$ is the index of each cell, drawn from a Gaussian random field with covariance matrix $\textbf{C}$.
   We can represent any linear function of the field as $\vec{u}\cdot\vec{f} = d$.
   Modifications of the initial conditions can be found by updating $\vec{f} \to \vec{f}'$ to satisfy the requirement
   \begin{equation}
      \vec{u}\cdot \vec{f'} = d',
      \label{eq:linear-constraint}
   \end{equation} where $\vec{f'}$ has the same power spectrum as $\vec{f}$ and $d'$ is an arbitrary value.
   In general, \cref{eq:linear-constraint} does not admit a unique solution.
   The genetic modification technique ensures unicity by further requiring $\vec{f'}$ to minimize the $\chi^2$ distance to the original field, \ie to minimize $(\vec{f'}-\vec{f})^\mathrm{T}\textbf{C}^{-1}(\vec{f'}-\vec{f})$.
   Note that $\vec{f}$ may be the density field $\vec{\delta}$ or the potential field $\vec{\phi}$.
}

In order to generate the initial conditions of the resimulations presented in \cref{sec:methods}, we implemented angular momentum modifications in \genetic{}.
{Given \cref{eq:linear-constraint}, this requires computing the three vectors $\vec{u}_{i}$ (where $i=1,2,3$ for $x$, $y$ and $z$ directions respectively), such that $\vec{u}_i\cdot \vec{f} = {l}_{i,0}$, where $l_{i,0}$ is the $i$-th component of the \spAM in the initial conditions.}
For a given Lagrangian patch $\Gamma$, we can compute the \spAM in the initial conditions using linear theory, as detailed in \cref{sec:details-TTT}.
\Cref{eq:initial-spAM} shows in particular that the initial \spAM can be computed from the reduced potential field $\phi$.

\begin{figure}
   \centering
   \includegraphics[width=0.66\columnwidth]{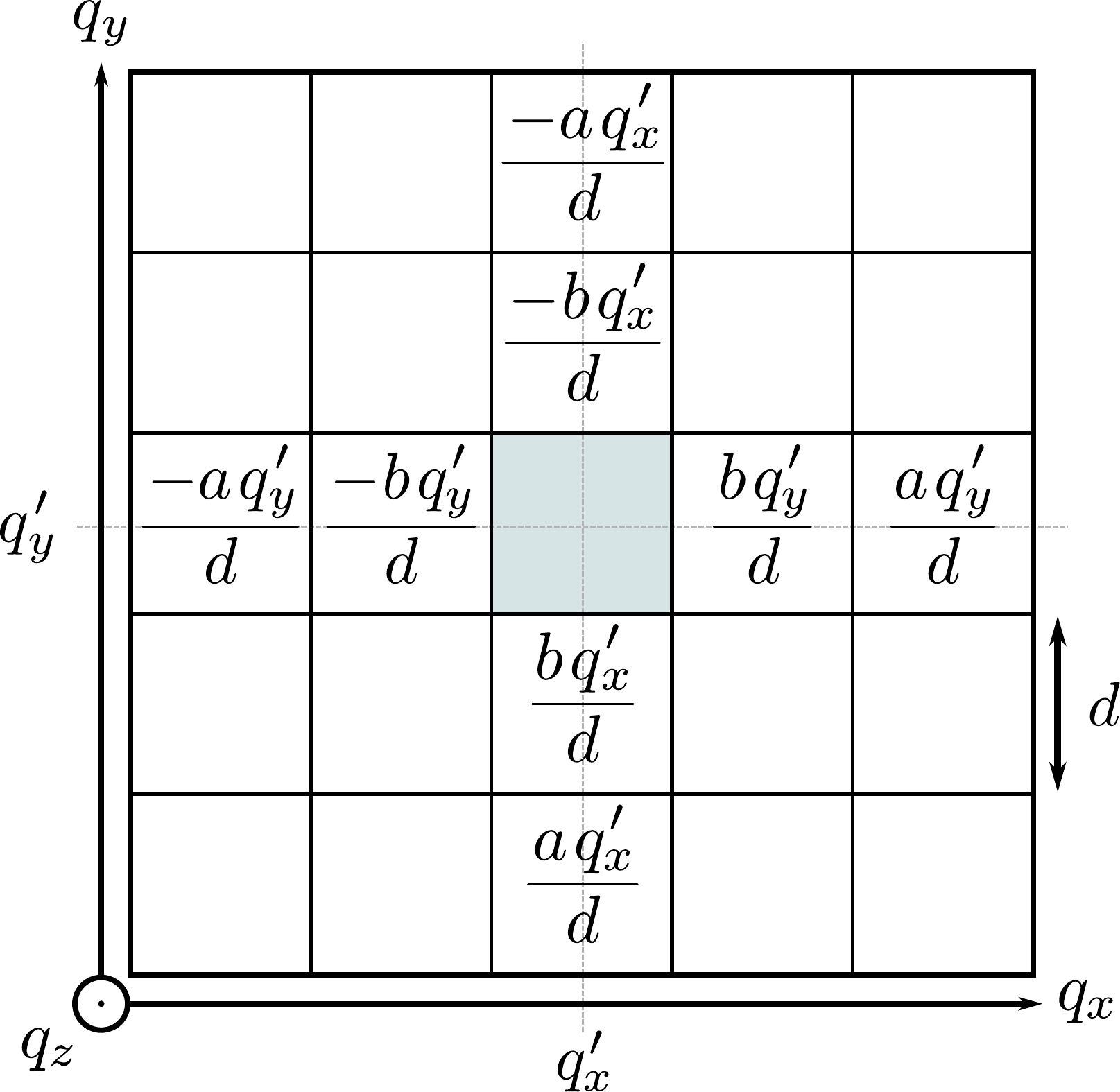}
   \caption{
     Sketch of the angular momentum operator in the $z$ direction. The operator acts on the reduced potential $\phi$.
     For each cell in the grid (the current one is highlighted in light blue, at position $q_x' = q_x-\bar{q_x}$ and $q_y' = q_y-\bar{q_y}$), we add the value shown on the sketch to the weight of the operator in the neighbouring cells.
     Here $a\equiv 1/12$ and $b\equiv -2/3$ are the coefficient of the fourth-order finite-difference scheme, and $d$ is the grid spacing.
     The action of the operator on the potential $\phi$ {yields for each cell $(q_x'\nabla_y{\phi} - q_y'\nabla_x{\phi})$ which, once summed over all cells in $\Gamma$, yields \cref{eq:AM-x-direction}}.
     This can be easily generalized to the two other directions.
   }\label{fig:AM-operator-sketch}
 \end{figure}

To do so, our algorithm first computes the reduced potential $\phi$ on a regular grid, from the over-density {$\vec{\delta}$}. We achieve this by solving the Poisson equation in Fourier space.
{%
   Without loss of generality, let us focus on the particular case of the \AM in the $z$ direction. Its value on a regular grid can be derived from \cref{eq:initial-spAM} as
   \begin{equation}
      l_{z,0} \propto \sum_{i}\left[ (q_{x}^{(i)} - \bar{q}_x) \nabla_y\phi^{(i)} - (q_{y}^{(i)} - \bar{q}_y) \nabla_x\phi^{(i)}\right],
      \label{eq:AM-x-direction}
   \end{equation}
   where $i$ runs over all the cells in $\Gamma$, $\bar{\vec{q}}$ is the center of mass
   and we obtain the value of $\nabla_x \phi^{(i)}$ using a fourth-order finite difference method.
   We then find the linear operator $\vec{u}_z$ such that $\vec{u}_z\cdot\vec{\phi} = l_{z,0}$, where $\vec{\phi}$ is a vector containing the values of the potential in all cells of $\Gamma$.
   To do so, we loop over all cells in $\Gamma$ and update the value of $\vec{u}_z$ as illustrated on \cref{fig:AM-operator-sketch}.
   Once computed, the value of $\vec{u}_z$ can readily be used to modify the angular momentum using the code \genetic.
}

We have used this algorithm to generate the initial conditions in our suite of resimulations presented in \cref{sec:methods}.

\label{sec:details-TTT}

\section{Details of tidal torque theory}

In \cref{sec:results} we compared the prediction from \TTT, \cref{eq:predictor-TTT}, to those relying on resimulations with modified initial conditions, \cref{eq:predictor}.
{For the sake of completeness,} we provide here a short {re}derivation of \TTT{}{, following broadly the lines of \cite{white_AngularMomentumGrowth_1984}}.
To do so, let us first consider a Eulerian patch $\gamma$. Its angular momentum about the centre of mass, located at comoving position $\bar{\xx}$ reads
\begin{equation}
  \vec{L} = a^5 \int_\gamma \dd[3]{x} \rho (\xx-\bar{\xx}) \cross \vec{v},
\end{equation}
where $\xx$ is the comoving position, $\vec{v}$ is the peculiar velocity, $a$ is the expansion factor and $\rho$ is the density.
Using the fact that $a^3\rho = a_0^3\bar{\rho_0}(1+\delta)$, with $\delta$ the overdensity, $a_0$ the initial expansion factor and $\rho_0$ the initial expansion density, we can express the \AM as
\begin{equation}
   \vec{L} = a^2\bar{\rho_0}a_0^3\int_\gamma\dd[3]{x}(1+\delta) (\xx-\bar{\xx}) \cross \vec{v}.
  \label{eq:AM-definition0}
\end{equation}
We can relate the comoving Eulerian position and velocity of a particle to its initial Lagrangian position plus a displacement,
\begin{equation}
  \xx(\qq, t) = \qq + \vec{S}(\qq, t), \quad \vec{v}(\qq, t) = \pdv{\vec{S}(\qq, t)}{t}.
\end{equation}
When the displacement is small, the mapping is reversible and the flow is laminar.
Under this assumption, we can write the mass conservation equation as
\begin{equation}
  \dd[3]{x} (1+\delta) = \dd[3]{q},
\end{equation}
so that \cref{eq:AM-definition0} becomes
\begin{equation}
  \vec{L} = a^2 \bar{\rho_0}a_0^3\int_\Gamma \dd[3]{q} (\xx - \bar{\xx})\cross \vec{v},
  \label{eq:AM-lagrangian}
\end{equation}
where $\Gamma$ is the Lagrangian region corresponding to the Eulerian region $\gamma$.
Note that this equation is exact in the laminar regime.

We can further simplify \cref{eq:AM-lagrangian} using the Zel'dovich approximation \citep{zeldovich_GravitationalInstabilityApproximate_1970a}.
In particular, we will assume that the proportionality factor between the velocity potential and the gravitational potential $\phi(\qq, t)$, which holds in the linear regime for the growing
mode of density fluctuations $\delta \propto D(t)$, can be extended into the mildly non-linear regime.
This allow us to write in the mildly non-linear regime
\begin{equation}
  \vec{S}(\qq, t) = - D(t)\grad{\phi(\qq)},
  \label{eq:displacement}
\end{equation}
where $\phi$ is the solution to the (reduced) Poisson equation $\laplacian \phi = \delta$.
Plugging \cref{eq:displacement} into \cref{eq:AM-lagrangian}, we finally obtain an expression for the angular momentum
\begin{equation}
   \vec{L} = -a^2 \dot{D}\bar{\rho_0}a_0^3 \int_\Gamma \dd[3]{q} (\qq - \bar{\qq}) \cross \grad{\phi},
   \label{eq:AM-zeldovich}
\end{equation}
where $\dot{D} \equiv \pdv*{D}{t}$.
We can obtain the \spAM $\vec{l}$ by diving \cref{eq:AM-zeldovich} by the mass $\rho_0 a_0^3V_\Gamma$ enclosed in the region with Lagrangian volume $V_\Gamma=\int_\Gamma\dd[3]q$, to obtain
\begin{equation}
   \vec{l}(t) = -\frac{a^2\dot{D}}{V_\Gamma} \int_\Gamma \dd[3]{q} (\qq - \bar{\qq}) \cross \grad{\phi}.
\end{equation}
We can split this into a time-dependent part and a space-dependent part as
\begin{equation}
   \vec{l}(t) = \vec{l}_0 f_\mathrm{TTT}(t).
   \label{eq:TTT-L}
\end{equation}
The initial value of the \spAM{} $\vec{l}_0$ is given by
\begin{equation}
   \vec{l}_0 \equiv -\frac{a^2_0\dot{D}_0}{V_\Gamma} \int_\Gamma \dd[3]{q} (\qq - \bar{\qq}) \cross \grad{\phi},
   \label{eq:initial-spAM}
\end{equation}
and the \AM growth rate is given by
\begin{equation}
   f_\mathrm{TTT}(t) \equiv \frac{a(t)^2\dot{D}(t)}{a_0^2 \dot{D}_0}.
\end{equation}
Note that \cite{white_AngularMomentumGrowth_1984} suggested to further simplify \cref{eq:initial-spAM} using a Taylor development of the potential about the center of the patch.
This allows to express the initial \spAM in terms of the density and potential fields only, independently of $\Gamma$, but at the cost of a decreased accuracy.
In this paper, we compared predictions of the \AM with known initial values, and therefore did not follow this route.
{The expression of \protect\cref{eq:initial-spAM} resembles the `spin from primordial inner motions' concept \protect\citep{neyrinck_HaloSpinPrimordial_2020}. In this formulation, the initial angular momentum is expressed as a function of the potential gradient, or equivalently of the peculiar velocity field in the Zeldovich approximation.}

In \cref{sec:methods}, \TTT's prediction (Eq.~\ref{eq:predictor-TTT}) was derived by taking the norm of both sides of \cref{eq:TTT-L}.
The form of the latter equation also suggests ansatz \eqref{eq:predictor}, where we replaced $f_\mathrm{TTT}$ with the measured growth rate in simulations, $l_\mathrm{ref}(z) / l_{\mathrm{ref},0}$.

\bsp  %
\label{lastpage}
\end{document}